\documentclass[twocolumn,showpacs,preprintnumbers,amsmath,amssymb]{revtex4}

\usepackage{graphicx}
\usepackage{dcolumn}
\usepackage{bm}
\usepackage{epsfig}

\begin{document}


\title{A new way to evaluate x-ray Brillouin scattering data}

\author{U. Buchenau}
 \email{buchenau-juelich@t-online.de}
\affiliation{%
J\"ulich Center for Neutron Science, Forschungszentrum J\"ulich\\
Postfach 1913, D--52425 J\"ulich, Federal Republic of Germany
}%
\date{February 28, 2014}

\begin{abstract}
Making use of the classical second moment sum rule, it is possible to convert a series of constant-$Q$ x-ray Brillouin scattering scans ($Q$ momentum transfer) into a series of constant frequency scans over the measured $Q$ range. The method is applied to literature results for the phonon dispersion in liquid vitreous silica and in glassy polybutadiene. It turns out that the constant frequency scans are again well fitted by the damped harmonic oscillator function, but now in terms of a $Q$-independent phonon damping depending exclusively on the frequency. At low frequency, the sound velocity and the damping of both evaluations agree, but at higher frequencies one gets significant differences. The results in silica suggest a new interpretation of x-ray Brillouin data in terms of a strong mixing of longitudinal and transverse phonons toward higher frequencies. The results in polybutadiene enlighten the crossover from Brillouin to Umklapp scattering.
\end{abstract}

\maketitle

Our knowledge of the sound waves at and above the boson peak in glasses is to a large part due to x-ray Brillouin scattering measurements \cite{ruffle1,ruffle2,mascio,monaco,baldi}, which allow to see the longitudinal part of the sound wave motion in the frequency range between 2 and 20 meV. The experimental arrangement makes scans of $S(Q,\omega)$ at constant momentum transfer $Q$ much easier than constant-$\omega$ scans \cite{mascio}. It is usual to fit such a constant-$Q$ scan in terms of the damped harmonic oscillator function, the so-called DHO
\begin{equation}\label{dho}
	\frac{S(Q,\omega)}{S(Q)}=f_Q\delta(\omega)+\frac{1-f_Q}{\pi}\ \frac{\Omega_Q^2\Gamma_Q}{(\omega^2-\Omega_Q^2)^2+\omega^2\Gamma_Q^2}.
\end{equation}
Here symbols with the index $Q$ depend on the momentum transfer $Q$, but {\it not} on the frequency $\omega$. $\Omega_Q$ is the sound wave frequency, which defines the sound velocity $c_Q=\Omega_Q/Q$ at this $Q$; $\Gamma_Q$ is the damping of the sound wave, and $f_Q$ is the elastic (in liquids quasielastic) fraction of the scattering at this $Q$. 

The weak point of this evaluation is the following: The strong damping which one fits to the sound waves above the boson peak is not a real physical damping of the vibrations at the sound wave frequency. Instead, it reflects a deviation of the eigenvectors from a perfect sine function in space. Thus, it is not a damping for all frequencies at fixed $Q$, as supposed by eq. (\ref{dho}), but rather a distribution of sound wave vectors around an average one at the given frequency. It is a property of the frequency window rather than a property of the momentum transfer window. In fact, this weak point can be directly seen at larger $Q$, where the DHO fit has too much intensity close to the elastic line \cite{ruffle1}.

On the other hand, at most points in the relevant $(Q,\omega)$-space, the DHO manages to fit the data very well. Thus it certainly supplies a good parameter set for the description of $S(Q,\omega)$. The question is only whether the parameters are indeed meaningful.

Fortunately, it is easy to translate a set of DHO measurements at a series of different $Q$ into the set of constant-$\omega$ scans which one would like to have. One notes first that for a DHO $(1-f_Q)S(Q)$ is fixed to the value
\begin{equation}
	(1-f_Q)S(Q)=\frac{k_BT}{Mc_Q^2}
\end{equation}
by the classical second moment sum rule \cite{donald}
\begin{equation}
	\int_{-\infty}^\infty\omega^2S(Q,\omega d\omega=\frac{k_BTQ^2}{M},
\end{equation}
where $M$ is the average atomic mass.

With this equation, one can calculate a constant-$\omega$ scan of $S(Q,\omega)$ for any $\omega$ in absolute units, each DHO-scan supplying a point at its $Q$-value. The result is best plotted in terms of the dimensionless dynamical structure factor $F_\omega(Q)$ defined by
\begin{equation}\label{fomq}
F_\omega(Q)=\frac{M\omega^3S(Q,\omega)}{k_BTQ^2},	
\end{equation}
which in terms of the DHO parameters is given by
\begin{equation}
	F_\omega(Q)=\frac{1}{\pi}\frac{\Gamma_Q\omega^3}{(\omega^2-\Omega_Q^2)^2+\omega^2\Gamma_Q^2}.
\end{equation}

Fig. 1 shows scans of $F_\omega(Q)$ for the beautiful Brillouin x-ray data of Baldi, Giordano, Monaco and Ruta \cite{baldi} in vitreous silica at 1620 K. In such scans, the broadening of the phonons in $Q$ can be directly seen in their own frequency window.

\begin{figure}[b]
\hspace{-0cm} \vspace{0cm} \epsfig{file=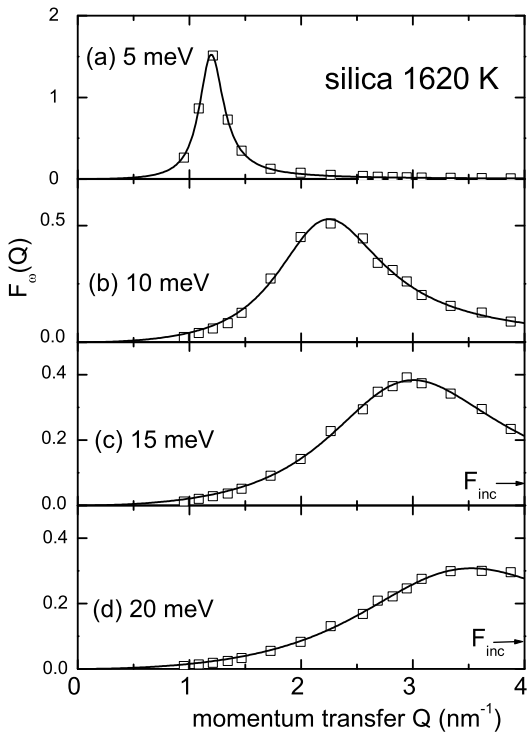,width=7 cm,angle=0} \vspace{0cm} \caption{Constant-$\omega$ scans of the dynamic structure factor $F_\omega(Q)$ calculated from the x-ray Brillouin scattering data of Baldi et al \cite{baldi} in vitreous silica at 1620 K. The lines are fits in terms of eq. (\ref{dhom}).}
\end{figure}

In order to fit these data, one uses the dynamic structure factor of a damped longitudinal phonon \cite{donald,boon}
\begin{equation}\label{dhom}
F_\omega(Q)=\frac{f_\omega}{\pi}\frac{(\Gamma_\omega/\omega)Q^2Q_B^2}{(Q^2-Q_B^2)^2+(\Gamma_\omega/\omega)^2Q^4}
\end{equation}
which is again the DHO, but now with parameters which no longer depend on $Q$. Instead, they depend on $\omega$ as they should (remember that the phonon broadening is a property of the phonon in a given frequency window, because it corresponds to a broadening in wavevector space and {\it not} in frequency). The Brillouin wavevector $Q_B$ defines the frequency-dependent longitudinal sound velocity $c_\omega=\omega/Q_B$.

As seen from Fig. 1, one gets excellent fits with eq. (\ref{dhom}). However, one can no longer reckon with the normalization property of the second moment sum rule. Therefore, one needs not only the two parameters $c_\omega$ and $\Gamma_\omega$, but an additional normalization factor $f_\omega$ as well. $f_\omega$ is found to decrease from 1 at low frequency to 0.75 at 25 meV, showing a slow disappearance of the acoustic correlation toward higher frequencies.

\begin{figure}[b]
\hspace{-0cm} \vspace{0cm} \epsfig{file=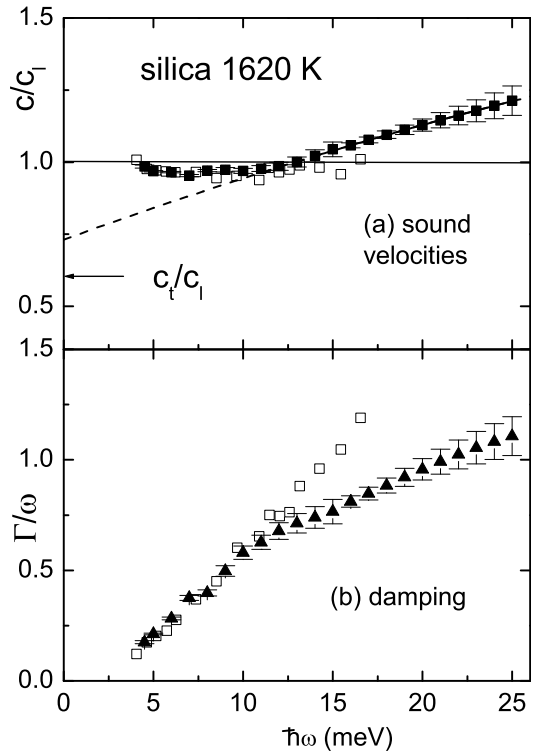,width=7 cm,angle=0} \vspace{0cm} \caption{Comparison of the DHO results of Baldi et al \cite{baldi} (open symbols) with those fitted here (full symbols) for (a) the sound velocity in units of the longitudinal light scattering Brillouin sound velocity $c_l$ (b) for the damping ratio $\Gamma/\omega$.}
\end{figure}

The fitting in terms of the three parameters $f_\omega$, $c_\omega$ and $\Gamma_\omega$ works so well that one even gets reliable numbers when the phonon peak begins to leave the measured $Q$-range, as in Fig. 1 (e). The fitted values for $c_\omega$ and $\Gamma_\omega$ are plotted in Fig. 2. Below 10 meV, the exchange of scattering laws does not bring anything new: The new parameters $c_\omega$ and $\Gamma_\omega$ agree within experimental error with the old $c_Q$ and $\Gamma_Q$ determined directly from the DHO scans. Nevertheless, even at these low frequencies the exercise is useful: It is not enough to know, one must also know that one can trust what one knows.

Between 10 and 25 meV, the new evaluation reveals not only a smaller damping of the phonons, but also a marked increase of the sound velocity with increasing frequency. A similar result has been obtained in the conventional way in glycerol \cite{monaco}, though there the initial decrease of the sound velocity is more pronounced than in silica and the subsequent hardening is less pronounced. This will also be seen in our second example, polybutadiene. 

The initial decrease and the subsequent hardening of the sound velocity can be understood by considering a mechanism which so far has not been taken into account, namely the possible interaction between longitudinal and transverse sound waves. The reason for the damping may be controversial, but whatever it is, the damping mechanism affects both longitudinal and transverse phonons. Thus one must expect a coupled longitudinal-transverse mode as soon as $\Gamma_Q$ is of the order of the difference of the sound wave frequencies at the given $Q$ (or, what amounts to the same, the wavevector broadening becomes of the order of the wavevector difference of longitudinal and transverse phonons at the given frequency $\omega$). This coupled mode will be expected to consist to one third of longitudinal and two thirds of transverse modes, because there are two transverse modes.

\begin{figure}[b]
\hspace{-0cm} \vspace{0cm} \epsfig{file=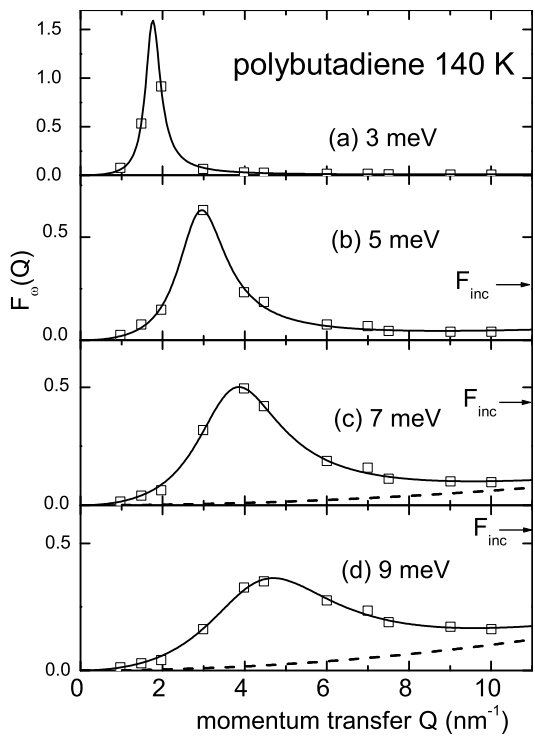,width=7 cm,angle=0} \vspace{0cm} \caption{Constant-$\omega$ scans of the dynamic structure factor $F_\omega(Q)$ calculated from the x-ray Brillouin scattering data of Fioretto et al \cite{fio} in polybutadiene at 140 K. The continuous lines are fits in terms of eq. (\ref{dhom}) with an added Umklapp term (the dashed lines in (c) and (d)) as explained in the text.}
\end{figure}

In fact, if one extrapolates the measured values for the high frequency sound velocity in Fig. 2 (a) back to the frequency zero, one arrives at the point corresponding to a sound velocity $c_l/3+2c_t/3$, the sound velocity which one expects for a mixed mode which is one third longitudinal and two thirds transverse.

This finding suggests the following interpretation of the data: There is a marked increase of the sound velocities with increasing frequency at all frequencies. However, at small frequency this increase is masked by the gradual transformation of the longitudinal sound waves into mixed ones, which leads to a concomitant increase of the wave vector of the mixed phonon in a Brillouin scattering experiment. In glycerol \cite{monaco}, this wavevector increase even overcompensates the hardening, leading to an apparent softening of the phonons with increasing frequency at the beginning of the dispersion curve.

\begin{figure}[b]
\hspace{-0cm} \vspace{0cm} \epsfig{file=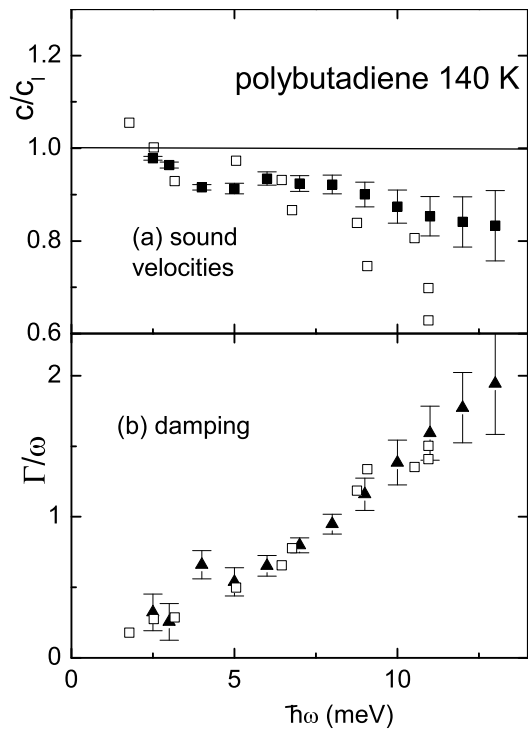,width=7 cm,angle=0} \vspace{0cm} \caption{Comparison of the DHO results of Fioretto et al \cite{fio} (open symbols) with those fitted here (full symbols) for (a) the sound velocity in units of the longitudinal light scattering Brillouin sound velocity $c_l$ (b) for the damping ratio $\Gamma/\omega$.}
\end{figure}

The fit of the silica data in Fig. 1 leaves an important open question: Why does one not see any indication of the Umklapp scattering, which dominates the dynamic structure factor at higher momentum transfer? Naturally, x-ray scattering is coherent scattering, but if we had incoherent scattering, then in the one-phonon approximation
\begin{equation}
	S_{inc}(Q,\omega)=\frac{k_BTQ^2}{2M}\frac{g(\omega)}{\omega^2},
\end{equation}
where $g(\omega)$ is the vibrational density of states. This leads to the constant dynamic structure factor $F_{inc}$
\begin{equation}
	F_{inc}=\frac{1}{2}\omega g(\omega)
\end{equation}
around which the coherent dynamic structure factor oscillates at higher $Q$.

For vitreous silica, $g(\omega)$ has been measured \cite{wischi} at 1673 K, close enough to the temperature of the x-ray Brillouin experiment \cite{baldi} of 1620 K to calculate a reliable $F_{inc}$. The values are shown by arrows in Fig. 1 (c) and (d), where they begin to be of the order of the measured ones. Why is it then possible to evaluate the data without any Umklapp contribution?

The question is answered by the evaluation of Brillouin x-ray data for polybutadiene \cite{fio} in Fig. 3. In this case, the data stretch to higher $Q$-values and the Umklapp scattering becomes indeed visible at higher momentum transfer in Fig. 3 (c) and (d). But even though one measures at more than twice the momentum transfer of Fig. 1, the Umklapp scattering is still a factor of six to seven weaker than the $F_{inc}$ calculated from a measurement \cite{gompbh} of $g(\omega)$ at 140 K.

The reason for this is the fact that the coherent scattering from any vibrational eigenmode must begin with $Q^4$, because the sum over the atomic displacements in the mode equals zero due to momentum conservation. This implies that the coherent dynamic structure factor $F_\omega(Q)$ begins with a $Q^2$-term. Obviously, in the momentum transfer range of Fig. 1 and Fig. 3, one is still in this initial region and can fit the Umklapp scattering with an additional $f_UQ^2$. This has been done for the data of Fig. 3. The dashed curves in Fig. 3 (c) and (d) show the Umklapp contribution alone.

Fig. 4 compares the resulting sound velocities and dampings with those obtained by the conventional evaluation. Again, the new evaluation provides more accurate data for the sound velocity, showing the initial softening and even a subsequent small hardening. The hardening, however, is barely seen, similar to the findings in glycerol \cite{monaco}.

From the two examples shown, it is obvious that one gets more (and more accurate) information from the new evaluation method proposed here, not only because it is better adapted to the physics, but also because it allows for a quantitative combination of several constant-$Q$ scans to calculate the dynamic structure factor on an absolute scale. For future experiments, it is naturally not necessary to fit with the DHO, because one can apply the second moment sum rule directly to the measured data. This should allow to extend the method beyond the Brillouin scattering range, where the data are no longer well fitted by the DHO.

To conclude, the classical second moment sum rule allows to calculate constant energy Brillouin scans from damped harmonic oscillator fits of constant-$Q$ scans. Applying the method to measurements in liquid vitreous silica, one finds a pronounced hardening of the Brillouin phonons at higher frequency, which suggests a radical change of previous interpretations. Instead of an essentially frequency-independent or even softening sound velocity, it now appears that one has to reckon with a marked hardening towards higher frequencies in the whole frequency range, a hardening which is only masked at the beginning by the mixing process of longitudinal and transverse phonons. The Umklapp scattering, not yet visible at the low momentum transfer of the vitreous silica scans, could be identified in measurements of polybutadiene at higher momentum transfer and was found to be surprisingly low in the Brillouin range.

\end{document}